\documentclass{iopart}
\usepackage[utf8]{inputenc}
\usepackage[T1]{fontenc}
\usepackage[english]{babel}
\usepackage{tikz}
\usetikzlibrary{shapes.geometric}
\usepackage{bbm}
\usepackage{listings, algpseudocode, amssymb}
\expandafter\let\csname equation*\endcsname\relax
\expandafter\let\csname endequation*\endcsname\relax
\usepackage{amsmath}
\newcommand{\pd}[2]{\frac{\partial#1}{\partial#2}}

\begin{document}
\title{Numerical Evolution from the Hamiltonian}
\author{José M L Amoreira$^1$ and Luís J M Amoreira$^2$}
\address{$^1$ Departamento de Física, Instituto Superior Técnico, Lisboa,
Portugal}
\address{$^2$ Departamento de Física, Universidade da Beira Interior, Covilhã,
Portugal}
\ead{amoreira@ubi.pt}
\begin{abstract}
  We propose a numerical method for approximate calculations of the time
  evolution of point particle systems given only the system's Hamiltonian
  function and initial conditions. The method both generates and solves the
  equations of motion numerically. For demonstration purposes, a working
  im\-ple\-men\-ta\-tion written in Python is described and applied
  to standard problems. The method may have some pedagogical merits but the
  numerical effort of generating the equations of motion makes it unsuitable for
  actual numerical solution of ``real'' problems with any but just a few degrees
  of freedom.
\end{abstract}

\submitto{\EJP}
%\maketitle
\section{Introduction}
%-----
% 1 As equações de Hamilton de um sistema de massas pontuais formam um sistema
% de N ODEs de 1ª ordem, com a forma 
% \dot x_i= G_i(t, \tilde x)
%-----
In the Hamiltonian formulation of classical mechanics~\cite{gol:1980, fw:2003},
the time evolution of physical systems is governed by Hamilton's equations
which, for point particle systems, take the form
\begin{align}\label{eq:heqs}
  \dot q_i&=\pd{H}{p_i}&
  \dot p_i&=-\pd{H}{q_i},&
  i&=1, 2, \ldots, N,
\end{align}
where $N$ is the number of degrees of freedom of the system (three times the
number of particles for three dimensional, unconstrained systems), $q_i$, $p_i$
are the canonical coordinates and conjugate momenta respectively, $H=H(t, q_i,
p_i)$ is the Hamiltonian function of the system (for systems with time
independent potentials and constraints, the Hamiltonian is simply the
mechanical energy function~\cite{fw:2003}) and dotted symbols denote their
total derivatives with respect to time. 
Renaming the canonical coordinates $q_i$ and momenta $p_i$ as
\begin{equation}\label{eq:xinot}
  \nu_i = 
  \begin{cases}
    q_i&\quad\text{if }i\leq N\\
    p_{i-N}&\quad\text{if }i> N
  \end{cases}
  \qquad i=1, \ldots, 2N,
\end{equation}
Hamilton's equations take a more unified form
\begin{equation}\label{eq:std_ode}
  \dot \nu_i=G_i(t, \nu_1, \nu_2, \ldots, \nu_{2N}), \quad i=1, \ldots, 2N
\end{equation}
where
\begin{equation}
  G_i(t, \nu)=
  \begin{cases}\displaystyle
    \pd{H}{\nu_{i+N}} ,&\quad\text{if }i\leq N\\[1em]
    \displaystyle
    -\pd{H}{\nu_{i-N}}, &\quad\text{if }i> N.
  \end{cases}
\end{equation}
This expression can be further shortened as a matrix product:
\begin{equation}\label{eq:rhsmf}
  G_i= \sum_j M_{i j}\pd{H}{\nu_j},\qquad
  \text{with }
  M=
  \begin{pmatrix}
    \mathbbm{O}&\mathbbm{I}\\
    -\mathbbm{I}&\mathbbm{O}
  \end{pmatrix},
\end{equation}
where $\mathbbm{O}$ and $\mathbbm{I}$ represent, respectively, the $N\times N$
zero and identity matrices.  Equation~\eqref{eq:std_ode} shows the Hamilton's
equations for a system of point particles with $N$ degrees of freedom as a
system of $2N$ first order ordinary differential equations (ODEs) on $2N$
variables $\nu_1, \nu_2, \ldots, \nu_{2N}$.

In all but a small handfull of very well known simple problems, this system of
ODEs has no analytical solutions and must be solved numerically.  Popular
computer library routines for solving ODEs\footnote{Like ODEPACK~\cite{odepack}
  for fortran, \texttt{odeint} \cite{odeint} in c++ boost libraries or
\texttt{solve\_ivp} in SciPy~\cite{scipy} for Python.} require the user to
supply subprograms to compute the right hand side functions $G_i$.  The ODE
solver invokes these subprograms to compute estimates for the values of the
unknowns $\nu_i$ for arbitrary time $t$, given their values at a particular
instant $t_0$.

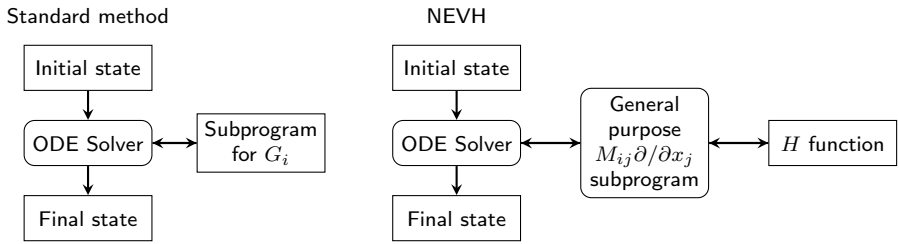
\begin{figure}[htb]
  \centering
    \begin{tikzpicture}[node distance=1.0cm, font=\sffamily,scale=0.7]
      \footnotesize
      \tikzstyle{libcode} = [rectangle,
                             rounded corners,
                             minimum width=1.5cm,
                             text width=1.5cm,
                             minimum height=0.6cm,
                             text centered,
                             draw=black,
                             %fill=orange!20
                             ]
      \tikzstyle{usercode} = [rectangle,
                             %rounded corners,
                             minimum width=1.5cm,
                             minimum height=0.6cm,
                             text centered,
                             draw=black,
                             text width=1.5cm,
                             %fill=green!20
                             ]
      \tikzstyle{arrow} = [thick,->,>=stealth]

      \begin{scope}[xshift=-3.5cm]
        \node at (0,1) {Standard method};
        \node (input) [usercode] {Initial state};
        \node (odesolver) [libcode, below of=input] {ODE Solver};
        \node (output) [usercode, below of=odesolver] {Final state};
        \node (rhs) [usercode, right of=odesolver, node distance=2.3cm]
          {Subprogram for $G_i$};
        \draw [arrow] (input) -- (odesolver);
        \draw [arrow] (odesolver) -- (output);
        \draw [arrow] (odesolver) -- (rhs);
        \draw [arrow]  (rhs) -- (odesolver);
      \end{scope}

      \begin{scope}[xshift=3.5cm]
        \node at (0,1) {NEVH};
        \node (input) [usercode] {Initial state};
        \node (odesolver) [libcode, below of=input] {ODE Solver};
        \node (output) [usercode, below of=odesolver] {Final state};
        \draw [arrow] (input) -- (odesolver);
        \draw [arrow] (odesolver) -- (output);
        \node (rhs) [libcode,  right of=odesolver,
                     node distance=2.5cm]
                     {General purpose\\ $M_{ij}%\,
                     \partial/\partial x_j$ subprogram};
        \draw [arrow] (odesolver) -- (rhs);
        \draw [arrow]  (rhs) -- (odesolver);
        \node (h) [usercode, right of=rhs, node distance=2.5cm] {$H$ function};
        \draw [arrow] (rhs) -- (h);
        \draw [arrow] (h) -- (rhs);
      \end{scope}
    \end{tikzpicture}
  \caption{Flowcharts for the standard and the proposed methods. Round cornered
    boxes represent library or general purpose code; right angled boxes contain
    code for the specific calculation at hand, which the user must
  supply.\label{fig:a}}
\end{figure}

In this work, we propose an alternative numerical method (which we will refer to
as \emph{Numerical Evolution from the Hamiltonian,} or NEVH) where the user only
needs to write code for the Hamiltonian function of the system, leaving its
partial derivatives to be computed numerically by a general purpose subprogram.
Figure~\ref{fig:a} displays flowcharts for both methods.

This method is computationally more expensive than the standard approach, since
at least two evaluations of the Hamiltonian function are needed to compute each
of its partial derivatives at any moment in time, and so it really is not
suitable for the analysis of complex systems with more than just a few coupled
degrees of freedom. However, for simple systems it is very practical
and, anyway, it is an interesting approach in its own, which we haven't seen
exposed elsewhere.

%-------------------------------------------------------------------------------
\section{Numerical procedure and a simple implementation in Python}
The ``Generic purpose $M_{ij}\partial/\partial x_j$ subprogram'' in the
flowchart in Figure~\ref{fig:a} can be easily implemented in any modern computer
language. Using a central difference formula for the partial derivatives, a
pseudocode version can be sketched as
\begin{algorithmic}[H]
  \Function{G}{$H, t, \nu$}
  \For{$1\leq i \leq 2N$}
  \State
  $\partial_iH\gets \frac{\displaystyle%
  H(t, \tilde\nu_i,\nu_i+\delta\nu_i)-
H(t, \tilde\nu_i,\nu_i-\delta\nu_i)}{\displaystyle2\delta\nu_i}
  $
  \EndFor\State
  \Return $\{\partial_{N+1}H, \partial_{N+2}H, \ldots, \partial_{2N}H,
  -\partial_1H, \partial_2H, \ldots, \partial_NH\}$
  \EndFunction
\end{algorithmic}
(In this pseudocode definition, the notation $\tilde\nu_i$ stands for all the
variables $\nu_1, \ldots, \nu_{2N}$, \emph{except} the $i$-th, $\nu_i$.)

We made a simple implementation of this method in Python, using the standard
packages NumPy~\cite{numpy:2011,numpy:2020} and SciPy~\cite{scipy}.  It
consists of a function that calls the Hamiltonian function supplied by
the user to evaluate, reorder and return its partial derivatives, computed using
second order central difference formulas.
%
%The step parameters $\delta\nu_i$ for
%the derivation formulas must be supplied by the user.
%This function is then
%given
%as an argument to SciPy's ODE solver \texttt{solve\_ivp}.
%
For greater simplicity of use, it is wrapped in a class (named \texttt{HGrad})
whose objects store, on creation, all the problem physical and numerical
details, like the Hamilton function and its parameters, the number of degrees of
freedom, the values of the discretization steps $\delta\nu_i$, etc.
\texttt{HGrad} objects are defined as callable with signature \texttt{objname(t,
$\nu_i$)}, returning a list containing the values of the rhs functions $G_i$ in
Hamilton's equations for given values of the dynamical variables of the
dynamical variables $\nu_i$ at given time \texttt{t}. In this way, the calling
syntax of \texttt{HGrad} objects is consistent with the interface of the ODE
solver we chose for our implementation (SciPy's function \texttt{solve\_ivp})
regarding the specification of the rhs $G_i$ functions, so that
\texttt{HGrad} objects can be used
just as if they were explicitly coded python functions.  The whole class
definition (including comments and Python doc strings) fits in a few dozen lines
of code and is very easy to use.  Our code is available for download as free
software at github~\cite{nevh:2020}.  The repository also includes a
\texttt{jupyter notebook} showcasing several example applications.

Listing~\ref{lst:1dlho} presents a particularly simple application example, the
one-dimensional harmonic oscillator. Most of the lines of code displayed deal
with general adminis\-trative details (initialization of physical and numerical
paramenters, definition of initial state, and so on), also needed for standard
numerical resolutions of Hamilton's equations. In Listing~\ref{lst:1dlho}, the
sections specific to our method are lines 6-8 (definition of the Hamiltonian
function), line 13 (initialization of the discretization steps for the
calculation of partial derivatives) and line 19 (definition of the
\texttt{HGrad} object \texttt{G}).

\begin{lstlisting}[caption= {Numerical solution of Hamiltonian's equations for
the one-dimensional linear harmonic oscillator. Note that the user only supplies
code for the system's Hamilton function in lines 7-9. The remaining code lines
define initial state, physical and numerical parameters.},
numbers=left, stepnumber=1, label={lst:1dlho}, language=python
]
from scipy.integrate import solve_ivp
import nevh
PI = 3.141592653589793

# Hamiltonian function
def H(t, s, k, m):
    x, p = s
    return 0.5 * k * x**2 + 0.5 * p**2 / m

# Initial state: off equilibrium position, at rest (x=1, p=0)
initial_state = [1.0, 0.0]
# Discretization step parameters for numerical derivatives
ds = [0.1, 0.1]
# Hamiltonian parameters. With k=4\pi^2, m=1, the period is 1
kc = 4 * PI**2
mc = 1.0

# Create the HGrad object, use scipy.solve_ivp to
G = nevh.HGrad(H, ds, k=kc, m=mc)

# solve Halmilton's eqs numerically from t=0 to t=3
trj = solve_ivp(G, [0.0, 3.0], initial_state)
\end{lstlisting}
Note that the \texttt{HGrad} object \texttt{G} is supplied to scipy's ODE solver
\texttt{solve\_ivp} just as if \texttt{G} was the name of a Python function for
evaluating the rhs of Hamilton's equations.

Scipy's ODE solver function \texttt{solve\_ivp} returns a composite structure
(named \texttt{trj} in Listing~\ref{lst:1dlho}) that stores information
on the numerical solution of the ODE, namely, sampled values of the
coordinates and momenta at different times in the interval \texttt{[tmin,
tmax]}.
%are stored at the arrays \texttt{trj.y[0]}, \texttt{trj.y[1]}, while the
%sampling times are stored in \texttt{trj.t}. The sampling frequency can be
%adjusted by the user, but here we just took the default behaviour of
%\texttt{solve\_ivp}, which sets that frequency accordind to precision
%requirements (that we also didn't bother to specify, because it wasn't necessary
%for convergence in this example). 

The values of the discretization steps used for the numerical computation of the
partial derivatives of the Hamiltonian are stored in array \texttt{ds} in
Listing~\ref{lst:1dlho} (at line 13).  In the example shown these values are
actually arbitrary, because the Hamiltonian of the linear harmonic oscilator is a
quadratic function of all the coordinates and conjugate momenta, and the central
difference formulas yield exact estimates of the derivatives of quadratic
functions. However, in other problems, the values of the discretizations steps
may need to be more carefully considered.

\begin{figure}[htb]
  \begin{center}
    \includegraphics[width=0.6\linewidth]{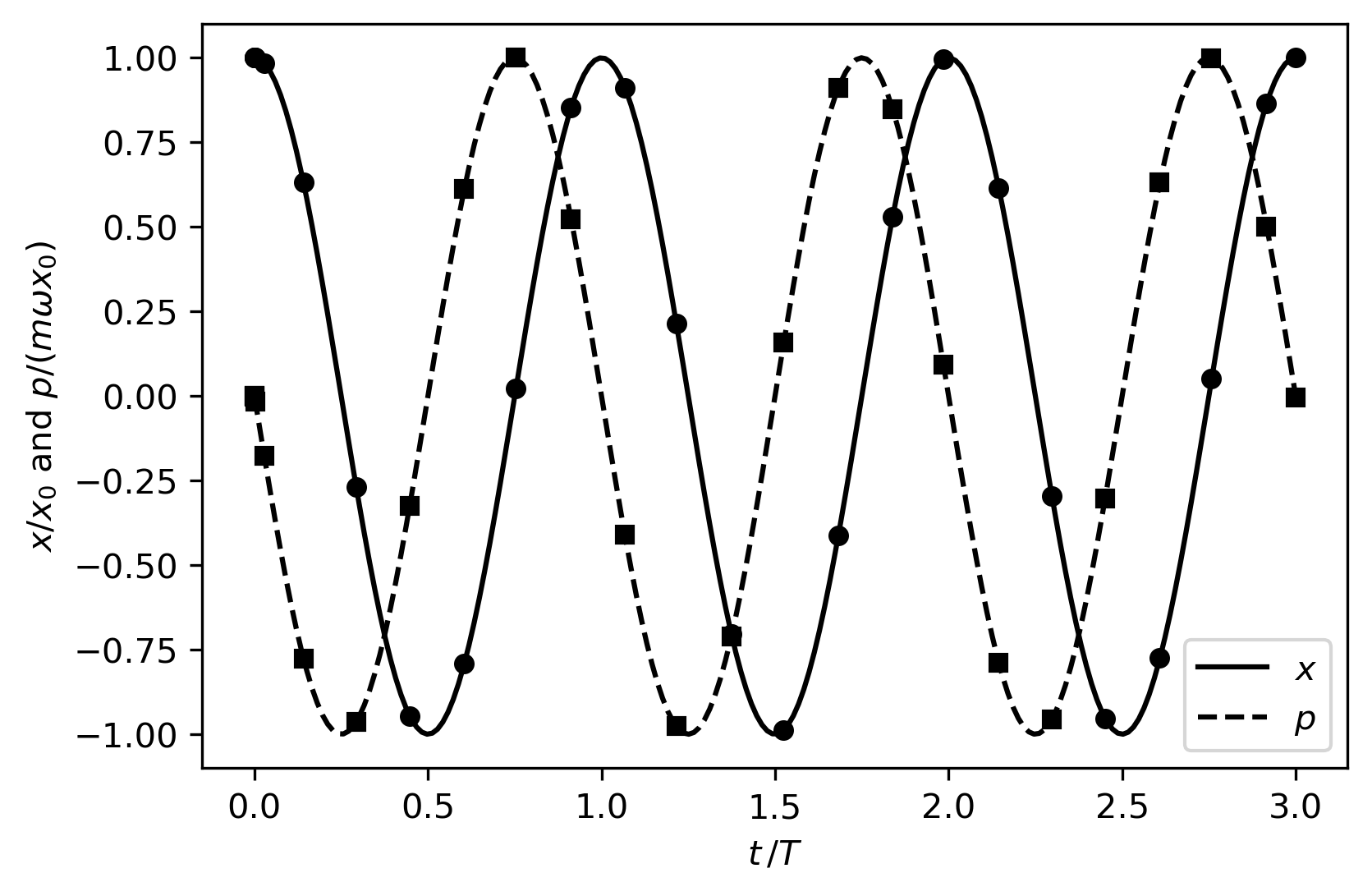}
  \end{center}
  \caption{Plot of position (\texttt{x}) and momentum (\texttt{p}) for the
  one-dimensional harmonic oscillator problem of Listing~\ref{lst:1dlho}.
  }\label{fig:1dlho}
\end{figure}

Figure~\ref{fig:1dlho} displays the values of the  coordinates and momenta of
the numerical solution obtained (circles and squares, respectively) in arbitrary
units, together with plots of the analytical solutions for the same problem
(continuous and dashed lines).
The correctness of the numerical approximation, at least for pedagogical
purposes only, is manifest.

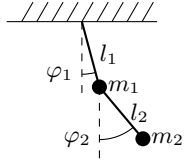
\begin{figure}[htb]
  {\centering
    \begin{tikzpicture}
      \small
      \draw (-1,0) -- (1,0);
      \foreach \x in {0.8, 0.6, 0.4, 0.2}
      {
        \draw (\x,0) -- +(60:0.3);
        \draw (-\x,0) -- +(60:0.3);
      }
      \draw (0,0) -- (60:0.3);
      \draw [thick] (0,0) -- (285:0.9) coordinate(a)
        node[midway, right]{$l_1$} 
        -- +(310:0.9) coordinate(b)
        node[midway,right] {$l_2$};
      \fill (a) circle(0.1) node [right]{$m_1$};
      \fill (b) circle(0.1) node [right]{$m_2$};
      \draw [dashed] (0,0) -- + (0,-1);
      \draw (285:0.7) arc (285:270:0.7)node [left] {$\varphi_1$};
      \draw [dashed] (a) --+ (0,-1);
      \path (a) -- +(310:0.7) coordinate(c);
      \draw (c) arc (310:270:0.7) node[left]{$\varphi_2$};
    \end{tikzpicture}
    \caption{\label{fig:b}Double pendulum.}
  }\par
\end{figure}

The preceding example is so simple that the main advantage of our method is not
really conveyed, since it's a trivial matter to derive and code the partial
derivatives of the harmonic oscillator Hamiltonian. Let us now consider a more
complex problem, the dynamics of the planar double pendulum. A planar double
pendulum is a set of two simple pendulums where the second is suspended from the
bob of the first (see Figure~\ref{fig:b}) and both are constrained to move in
the same vertical plane. This system has two obvious degrees of freedom, the
angles $\varphi_1$ and $\varphi_2$ which both pendulums define with the vertical
(see Figure~\ref{fig:b}).

For the case of equal masses $m$ and lengths $l$, the Hamiltonian
reads~\cite{bohm:2018}
\begin{equation}\label{eq:dph}
  H=\frac{1}{2ml^2}
  \frac{p_1^2 + p_2^2-2p_1p_2\cos(\varphi_1-\varphi_2)}%
    {1+\sin^2(\varphi_1-\varphi_2)}-
    mgl(2\cos\varphi_1+\cos\varphi_2),
\end{equation}
where the conjugate momenta $p_1$, $p_2$ are given by
\begin{align}
  p_1&=2ml^2\dot{\varphi_1}+ml^2\dot{\varphi_2}\cos(\varphi_1-\varphi_2)&
  p_2&=2ml^2\dot{\varphi_2}+ml^2\dot{\varphi_1}\cos(\varphi_1-\varphi_2)
\end{align}
Calculating the partial derivatives of this Hamiltonian relative to the angles
$\varphi_i$ and their conjugate momentum $p_i$ and programming the resulting
expressions is quite tedious and error-prone. Here, NEVH is markedly
simpler: the user just just needs to code the Hamiltonian~\eqref{eq:dph}, and
that's all. No need to derive and code complicated expressions for the right
hand side functions $G_i$ in Hamilton's equations~\eqref{eq:std_ode}.

\begin{figure}[htb]
  {\centering
    \includegraphics[width=0.8\linewidth]{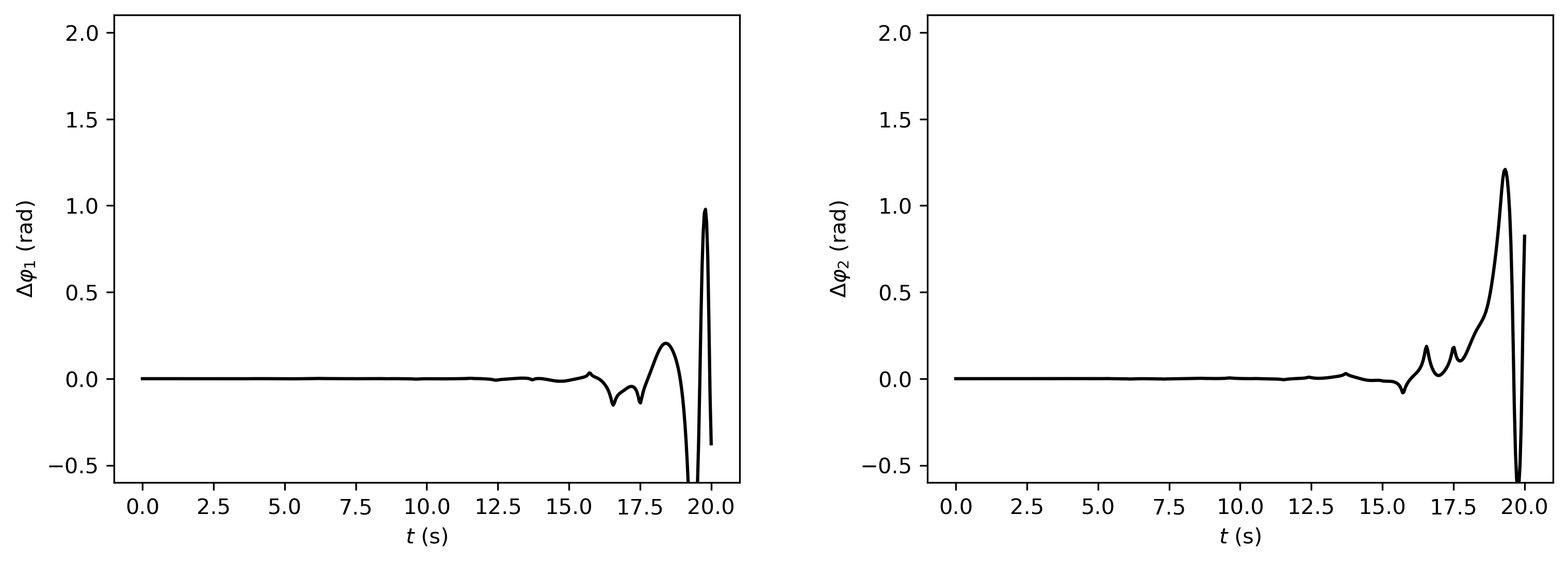}
    \caption{\label{fig:c}Double pendulum. The plots show the diference between
    the angle values ($\varphi_1$ on the left, $\varphi_2$ to the right)
    computed with the standard and the NEVH method. Initial state is
    $\varphi_1=\pi/2$, $\varphi_2=0$, $p_1=p_2=0$. The ODE solver
    (\texttt{scipy.integrate.solve\_ivp}) is configured similarly for both
  methods.}
  }\par
\end{figure}

For the double pendulum, we implemented the two methods.  Both approaches yield
similar results. Figure~\ref{fig:c} shows plots of the difference between
results for the two angles $\varphi_1$ and $\varphi_2$ obtained using both
methods, for the case $m=1$\,kg, $l=1$\,m, $g=9.8$\,m/s$^2$. The initial
configuration is $\varphi_1=\pi/2,\ \varphi_2=0,\ p_1=p_2=0$.
Both methods initially yield essentially the same results. However, due to the
well known chaotic nature of the double pendulum dynamics, small discrepancies
grow up and sooner or later lead to completely different predicted evolutions of
the system.  Expectedly, this amplification of differences can be somewhat
delayed by choosing smaller discretization steps in the numeric calculation of
the partial derivatives of the Hamiltonian, thus increasing their accuracy.

In this example, the Hamiltonian function is five python lines long; the
definition of the $G_i$ functions is twice as long, taking 12 lines of code. 
It must also be said, however, that the NEVH code for this problem is around
four times slower than the program for the standard approach.

\section{Accuracy}
\begin{figure}[htb]
  \centering
    \includegraphics[width=0.6\linewidth]{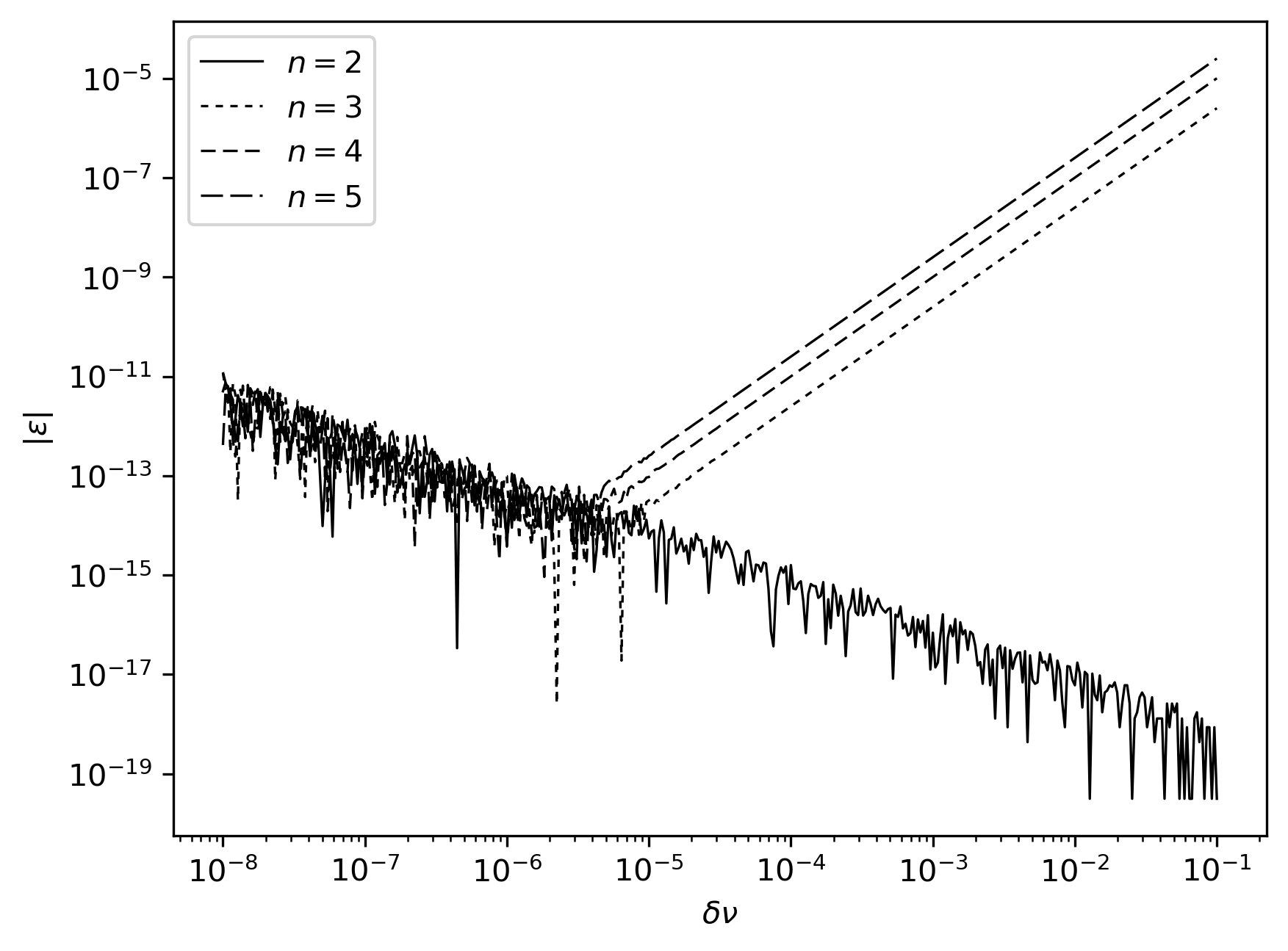}
  \par
  \caption{\label{fig:e}L2-norm of the difference between results obtained
    with the standard method and NEVH for a single time step (of duration
    $\delta t=0.001$, in arbitrary units) evolution of a point particle with
    hamiltonian \eqref{eq:errh} as a function of the discretizations steps
    $\delta\nu_1=\delta\nu_2\equiv\delta\nu$. The values $\alpha=m=1$ (arbitrary
    units) were used. The initial state was $x=1.0;\ p=0.0$.}
\end{figure}

The sources of numerical inaccuracy specific to the method here proposed are the
errors in the numerical evaluation of the partial derivatives of the
Hamiltonian. In our implementation, we used lowest order central difference
formulas, and so these errors are expected to be of second order on the
discretizations steps $\delta\nu_i$ (except for linear and quadratic functions,
for which central difference formulas yield exact results).
In order to check this, we compared results 
 for a single
time step evolution of a one-dimensional point partice subject to power law
potentials, 
with Hamiltonian function
\begin{equation}\label{eq:errh}
  H_{\alpha,n}(x,p)=\alpha x^n+\frac{1}{2m}p^2,
\end{equation}
for some real constant $\alpha$.
We took as measure of error the $L_2$-norm of the difference of final states
computed with NEVH and the standard approach.
The results are plotted in the graph displayed in Figure~\ref{fig:e} as 
functions of the discretization steps $\delta\nu_i$ (we
took $\delta\nu_1=\delta\nu_2\equiv\delta\nu$ in arbitrary units).

The graph displays the expected $O(\delta\nu^2)$ behaviour, for
$\delta\nu\gtrsim5\times10^{-6}$. For smaller values of $\delta\nu$, this tendency is
not perceived because floating point round-off errors become dominant (at the
floating point precision used in our calculation), hiding the contribution from
the calculation of the partial derivatives. For the case $n=2$, the numerical
derivation is exact and only round-off errors affect the results. These two
facts agree with our expectations, supporting the conviction that NEVH is a
well behaved, reliable method for the numerical solution of Hamilton's equations
for simple, low-dimensional point particle systems.

\section{Conclusion}
For numerical simulations of the dynamics of simple point particle systems
with few coupled degrees of freedom, NEVH is much simpler then the traditional
approach, because the user is spared the trouble of deriving expressions for the
partial derivatives of the Hamiltonian with respect to the system's coordinates
and momenta and of writing subprograms to compute them. The user only needs to
supply code for the system's Hamiltonian itself.

It is true that NEVH runs
slower than equivalent traditional approaches, because two or more evaluations
of tht Hamiltonian are needed to compute each of its partical derivatives. 
However, for simple systems, the time delay is hardly noticeable by the user.

Regarding accuracy, NEVH behaves as expected and (also as expected) does not
seem to introduce particular numerical instablities.

For these reasons, we find quite reasonable that NEVH shoud be considered in
numerical simulations of simple point particle systems, in particular for
pedagogical purposes.

%Note that we do not claim that this method is more efficient than the standard
%approach. Much to the contrary, in computationally intensive applications, the
%effort involved in the numerical calculation of the partial derivatives of the
%Hamiltonian (for which several computations of the Hamiltonian function itself
%are needed), at every instant in the simulation, slows the calculaton very
%significantly. What we do claim is that our method makes it very simple to set
%up a practical simulation of small mechanical systems, because the user is
%spared the trouble of deriving expressions for the partial derivatives of the
%Hamiltonian and writing subroutines that compute them, with the calling
%signature that the particular ODE solver expects.

\end{document}